# Pulse Shape Discrimination of Bulk and Very Bulk Events within HPGe Detectors

Jinfu Zhu, Tao Xue, Liangjun Wei, Jianmin Li

*Abstract–*The CDEX (China Dark matter Experiment) now deploys ~10 kg pPCGe (p-type Point Contact Germanium) detectors in CJPL (China Jinping Underground Laboratory). It aims to detect rare events such as dark matter and 0vbb (neutrinoless double beta decay). The discrimination of bulk and very bulk events are essential for improvements of the analysis threshold of dark matter. Very bulk events are generated near the p+ point surface of pPCGe, which are usually from radioactive materials of electronic devices. Due to different locations of charge collection, bulk and very bulk events have different pulse shape. This paper will present two linear PSD (Pulse Shape Discrimination) methods: CCM (Charge Comparison Method) and Fisher's LDA (Linear Discriminant Analysis), to realize the discrimination of bulk and very bulk events. The results show that FOMs (Figure of Merit) are 1.38±0.33 and 1.62±0.18 by CCM and Fisher's LDA respectively.

## I. INTRODUCTION

Dark matter direct detection [1] [2] and 0vbb (Neutrinoless double beta decay) of $^{76}$Ge observation [3] are very popular topics these years. Some experiments using HPGe (High Purity Germanium) detectors focus on dark matter candidates: WIMPs (Weakly Interacting Massive Particles) [4] [5]. These experiments require low radiation background and are generally deployed at deep underground laboratories, such CJPL (China Jinping Underground Laboratory) [6] and LNGS (Laboratori Nazionali del Gran Sasso) [7]. The CDEX (China Dark matter Experiment) [8] now uses ~10 kg pPCGe (p-type Point Contact Germanium) detectors in CJPL. Very bulk events occur near the p+ point surface of pPCGe, which are mainly from radioactive materials. Due to different locations of charge collection, bulk and very bulk events have different pulse shape. The discrimination of bulk and very bulk events is used to supress the background level [9]. This paper will use two linear PSD (Pulse Shape Discrimination) methods: CCM (Charge Comparison Method) [10] and Fisher's LDA (Linear Discriminant Analysis) [11]. They have been widely used in other detectors [12] [13] [14].

This paper is organized as follows. Section II describes bulk and very bulk events of CDEX HPGe detectors. Section III will introduce two linear PSD methods and their results. The conclusion and future work will be discussed in Section IV.

## II. BULK AND VERY BULK EVENTS

The structure of 1 kg pPCGe detector and simulation of the electric field [9] [15] are shown in Fig. 1 and Fig. 2. Events that occur in the bulk (called bulk events) have sufficient charge collection. The arrows in lines are motion directions of carriers. There exists a weak electric area (in colour blue) in the centre of detector. Very bulk events occur near the p-point face. Their charge collection doesn't need to go through the weak electric area so they has faster drift time.

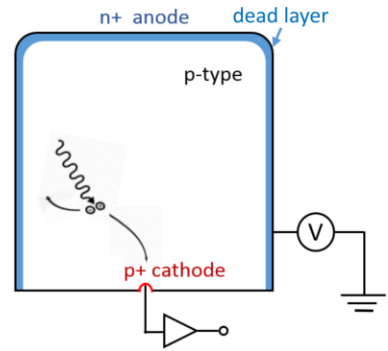

Fig. 1  The structure of 1 kg pPCGe detector.

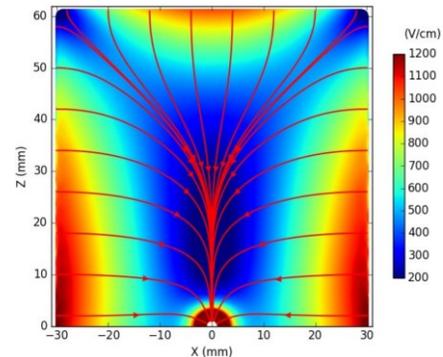

Fig. 2  Electric field simulation.

The discrimination of bulk and very bulk events are essential for improvements of the analysis threshold of dark matter. Very bulk events are generated near the p+ point surface of pPCGe, which are usually from radioactive materials of electronic devices. Due to different locations of charge collection, bulk and very bulk events have different pulse shape. Figure 3 presents typical bulk and very bulk events with 88 keV. The waveforms are from timing amplifiers (Canberra 2111) and digitized by 1 GSPS 12-bit

Manuscript received October 22nd, 2020. This work was supported by the National Key Research and Development Program of China (2017YFA0402202).

The authors are with the Department of Engineering Physics, Tsinghua University, and also with the Key Laboratory of Particle & Radiation Imaging (Tsinghua University), Ministry of Education, Beijing, 100084, China (Corresponding author is Tao Xue, e-mail: xuetao@mail.tsinghua.edu.cn).

ADC (Analog-to-Digital Converters). The black line shows the difference of two kind of pulses.

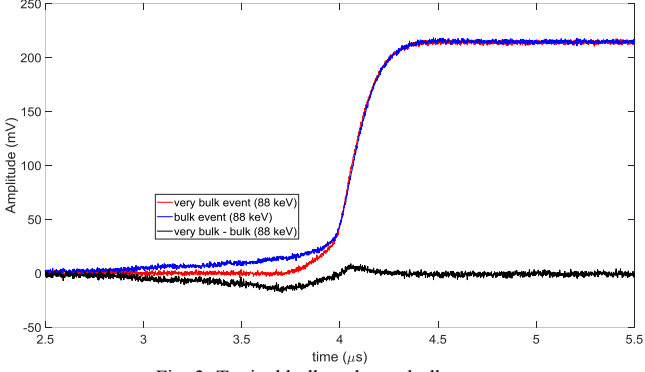

Fig. 3  Typical bulk and very bulk events.

## III. LINEAR PSD METHODS AND THEIR RESULTS

### A. CCM

Figure 4 depicts the normalized shapes of bulk and very bulk events. The CCM is used to realize PSD of two kind of events. The *CCM PSD ratio* is defined as equation (1). It uses two gate integrals of pulses.

$$PSD\ ratio = \frac{integral(later)}{integral(pre)} \quad (1)$$

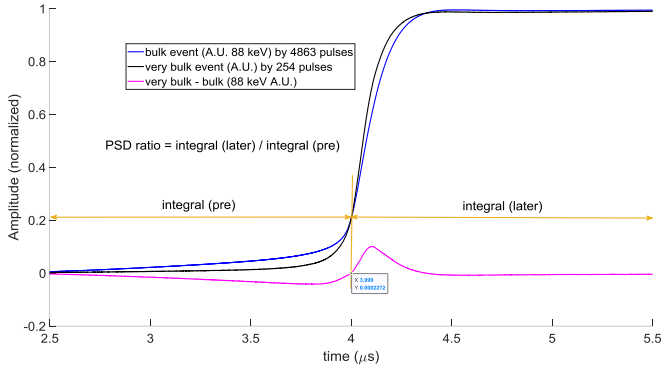

Fig. 4  The normalized shape of bulk and very bulk events.

### B. Fisher's LDA

Fisher's LDA is widely used in two-class classification problem. Two data sets of bulk and very bulk are as following:

$$\mathbf{D}_{bulk} = \{\mathbf{x}_{bulk_1}, \mathbf{x}_{bulk_2}, \cdots\}$$
$$\mathbf{D}_{vbulk} = \{\mathbf{x}_{vbulk_1}, \mathbf{x}_{vbulk_2}, \cdots\}$$

The variances of the two classes are shown as equation (2) and their separation is expressed as equation (3)

$$\mathbf{\Sigma}_{bulk} = \sum_{i=1}^{n}(\mathbf{x}_{bulk_i} - \mathbf{\mu}_{bulk})(\mathbf{x}_{bulk_i} - \mathbf{\mu}_{bulk})^T$$
$$\mathbf{\Sigma}_{vbulk} = \sum_{i=1}^{n}(\mathbf{x}_{vbulk_i} - \mathbf{\mu}_{vbulk})(\mathbf{x}_{vbulk_i} - \mathbf{\mu}_{vbulk})^T \quad (2)$$

$$J(\mathbf{w}) = \|\mathbf{w}^T\mathbf{\mu}_{bulk} - \mathbf{w}^T\mathbf{\mu}_{vbulk}\|_2^2 = \frac{\mathbf{w}^T(\mathbf{\mu}_{bulk} - \mathbf{\mu}_{vbulk})(\mathbf{\mu}_{bulk} - \mathbf{\mu}_{vbulk})^T\mathbf{w}}{\mathbf{w}^T(\mathbf{\Sigma}_{bulk} + \mathbf{\Sigma}_{vbulk})\mathbf{w}} \quad (3)$$

The optimal weight parameter $\mathbf{w}$ should meet $\mathbf{w} = \arg\max J(\mathbf{w})$. We let $\nabla J(\mathbf{w}) = 0$ and then obtain the solution as equation (4).

$$\mathbf{w} \propto (\mathbf{\Sigma}_{bulk} + \mathbf{\Sigma}_{vbulk})^{-1}(\mathbf{\mu}_{bulk} - \mathbf{\mu}_{vbulk}) \quad (4)$$

### C. Results

As shown in Fig. 5, the PSD *FOM* (Figure of Merit) is defined as equation (5), where $R_{VB}$ and $R_B$ are the mean value of very bulk and bulk events, and their variances are $\sigma_{VB}$ and $\sigma_B$ respectively.

$$FOM = \frac{|R_{VB} - R_B|}{2.355(\sigma_{R_{VB}} + \sigma_{R_B})} \quad (5)$$

The results show that FOMs (Figure of Merit) are 1.38±0.33 and 1.62±0.18 by CCM and Fisher's LDA respectively, as shown in Fig. 6 and Fig. 7.

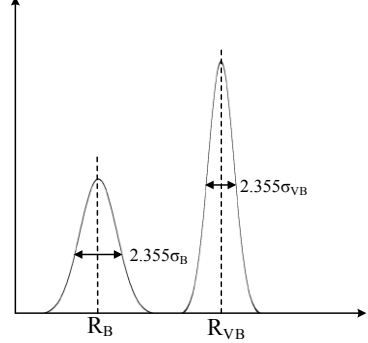

Fig. 5  The calculation of PSD FOM.

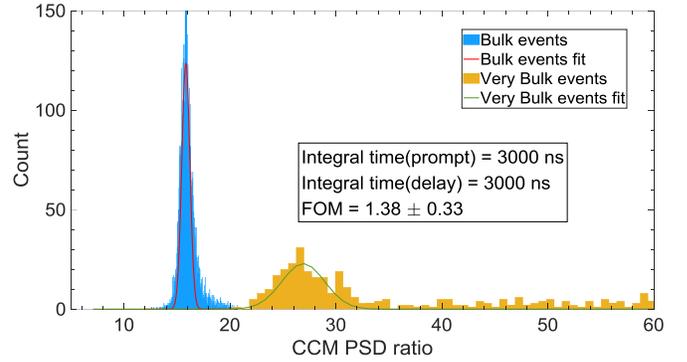

Fig. 6.  The PSD FOM by CCM

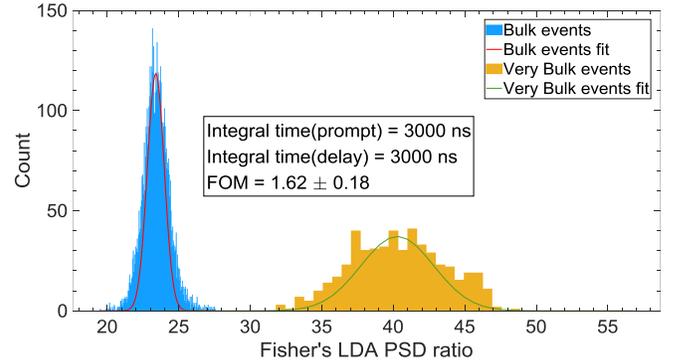

Fig.7.  The PSD FOM by Fisher's LDA

## IV. SUMMARY AND FUTURE WORK

This paper use and compare two linear PSD methods on bulk and very bulk events within CDEX HPGe detectors. The results show that Fisher's LDA has better performance than CCM. In the future, to further improve the PSD FOM of bulk and very bulk events, more quantitative analysis of discrimination performance and nonlinear methods will be researched.

ACKNOWLEDGEMENT

We would like to thank those who collaborated on the CDEX, and also thank Professor Hao Ma, Zhi Deng, Yulan Li, and Dr. Litao Yang for their support and various discussions over the years at the Tsinghua University DEP (Department of Engineering Physics).

We are grateful for the patient help of Yu Xue, Wenping Xue, and Jianfeng Zhang. They are seasoned full-stack hardware technologist with wealth experience of solder and rework in the electronics workshop at DEP.